\documentclass{article}
\usepackage{preamble}

\title{Evolving linguistic divergence \\ on polarizing social media}
\author{
Andres Karjus\textsuperscript{1,2,3}, Christine Cuskley\textsuperscript{4,5}\\
\small\textsuperscript{1} ERA Chair for Cultural Data Analytics, Tallinn University\\
\small\textsuperscript{2} School of Humanities, Tallinn University\\
\small\textsuperscript{3} Estonian Business School\\
\small\textsuperscript{4} English Literature, Language and Linguistics, Newcastle University\\
\small\textsuperscript{5} Centre for Behaviour and Evolution, Newcastle University
}
\date{\vspace{-1cm}}

\begin{document}
\maketitle
\begin{abstract}
\normalsize
Language change is influenced by many factors, but often starts from synchronic variation, where multiple linguistic patterns or forms coexist, or where different speech communities use language in increasingly different ways. Besides regional or economic reasons, communities may form and segregate based on political alignment. The latter, referred to as political polarization, is of growing societal concern across the world. 
Here we map and quantify linguistic divergence across the partisan left-right divide in the United States, using social media data. 
We develop a general methodology to delineate (social) media users by their political preference, based on which (potentially biased) news media accounts they do and do not follow on a given platform.
Our data consists of 1.5M short posts by 10k users (about 20M words) from the social media platform Twitter (now "X"). Delineating this sample involved mining the platform for the lists of followers (n=422M) of 72 large news media accounts.
We quantify divergence in topics of conversation and word frequencies, messaging sentiment, and lexical semantics of words and emoji. We find signs of linguistic divergence across all these aspects, especially in topics and themes of conversation, in line with previous research. 
While US American English remains largely intelligible within its large speech community, our findings point at areas where miscommunication may eventually arise given ongoing polarization and therefore potential linguistic divergence.
Our methodology --- combining data mining, lexicostatistics, machine learning, large language models and a systematic human annotation approach --- is largely language and platform agnostic. In other words, while we focus here on US political divides and US English, the same approach is applicable to other countries, languages, and social media platforms.

\end{abstract}

\section{Introduction}

All human languages change over time, as linguistic variants are discarded, innovated, and their meanings change. 
Most change likely stems from variation, whether geographical, cultural or social. Here we examine a division and source of variation intersecting these categories: political polarization. Social and political scientists have been increasingly concerned with the causes and alarming social effects of increasing media polarization and partisan segregation. 
While happening around the world, one country these effects appear to be particularly pronounced is the United States. The left-right divide has increased on the governmental level \parencite{andris_rise_2015} but also in everyday life, affecting where Americans choose to live \parencite[][]{mummolo_why_2017,brown_measurement_2021},  
how they raise their children \parencite{tyler_learning_2022}, %
how they deal with misinformation \parencite{petersen_need_2023,gonzalez-bailon_asymmetric_2023}, %
and which daily cultural and material products they consume \parencite{hetherington_prius_2018,rawlings_polarization_2022}.
In the information space, besides the growing divergence of news media \parencite{jurkowitz_us_2020,broockman_impacts_2022,muise_quantifying_2022}, 
polarization and segregation effects have been observed in
diverging public narratives about society and significant events \parencite{li_speaking_2017,demszky_analyzing_2019},
online knowledge curation \parencite{yang_polarization_2022},   %
as well as behavior on social media \parencite{adamic_political_2005,rathje_out-group_2021,mukerjee_political_2022,rasmussen_political_2022}.

Social media does not exist in an online vacuum, meaning it can affect lives in the real world.
For example, it has been shown that anti-refugee sentiment on Facebook predicts crimes against refugees in otherwise similar communities \parencite{muller_fanning_2021}, or that Twitter data like user network structure and message sentiment can predict results of future political elections \parencite[][]{jaidka_predicting_2019}.
Content personalization algorithms on social media platforms can (intentionally or not) amplify or diminish the visibility of political camps and messaging; \textcite{huszar_algorithmic_2022} show that US right-leaning officials and news sources enjoyed more amplification on Twitter compared to the left.

\subsection{Division and change}

Political polarization may also have an effect on the dynamics of language evolution and change, forming the basis for signals of in-group and out-group status \parencite{albertson_dog-whistle_2015}, with the potential to lead to more dramatic language speciation over time \parencite{andresen_languages_2016}. 
While American English varies naturally given the large geographic area and heterogeneous society it spans, it has been shown that there are growing linguistic differences that correlate with party affiliation in politicians \parencite{azarbonyad_words_2017,card_computational_2022,bhat_covert_2020,li_speaking_2017,wignell_twittering_2020}, 
as well as areas in the US with a strong left or right leaning 
\parencite{grieve_mapping_2018,louf_american_2023}.
If such divergence is or will become large enough to feasibly lead to misunderstanding in communication, then it can contribute to further polarization, potentially creating a ratchet effect which intensifies polarization.
Therefore, understanding emerging linguistic variation is a crucial component in understanding and eventually working towards easing sociopolitical polarization before it reaches a tipping point \parencite{macy_polarization_2021}.
While intervention experiments have shown it is possible to steer people away from misinformation and polarizing narratives
\parencite{pennycook_shifting_2021,balietti_reducing_2021,broockman_impacts_2022}, %
their efficiency is contingent on the ability of groups to communicate in the first place.

Some divergence in a given language may be attributed to natural linguistic drift or topical fluctuations \parencite{croft_explaining_2000,blythe_neutral_2012,karjus_quantifying_2020} taking different directions in groups with differing communicative needs \parencite{kemp_semantic_2018,karjus_conceptual_2021}, more so if they do not interact, and engage in different activities. 
Yet some lexical innovation and group-specific usage may be intentional.
One such example is that of the "dog whistle", as used in advertising or political communication: a word or phrase that is expected to mean one thing to the larger public, but carries an additional implicit meaning for a subset of the audience. For example, \textit{inner city} can mean "the area near the center of a city", but also signal an area with social problems or certain racial concentration. The finger gesture previously commonly meaning "okay" or "all good" has been appropriated by the far-right  \parencite[see][]{albertson_dog-whistle_2015,khoo_code_2017,bhat_covert_2020}. Such expressions can be used to circumvent censorship and moderation and convey messages that would be otherwise deemed unfit for publication, including hate speech.

\subsection{Online media as a data source}

In this contribution, we map and quantify linguistic divergence between the left-right political divide --- focusing on American English and lexical and semantic variation --- using a corpus of posts mined from the social media platform Twitter (at the time of writing, Twitter is in the process of being renamed to "X", but is still operational at www.twitter.com). The data was collected between between February and September 2021.
Twitter data --- while subject to a number of issues discussed below --- has been shown to be useful for mapping lexical variation and innovation and other socio-cultural processes \parencite{donoso_dialectometric_2017,dzogang_diurnal_2018,grieve_mapping_2018,bhat_covert_2020,robertson_emoji_2020,ananthasubramaniam_networks_2022,alshaabi_storywrangler_2021} 
and analyzing polarization dynamics \parencite{chen_polarization_2021,an_visualizing_2012,rathje_out-group_2021}. 
Studies of linguistic divergence between political divides have often focused on politicians and activists \parencite{adamic_political_2005,li_speaking_2017,gentzkow_measuring_2019}. 
Here we are interested in everyday language by regular speakers, to the extent it can be inferred from social media.

The variation we seek to quantify in this contribution is of course just one dimension of linguistic variation within a given language. American English, like other languages, also varies across geography (referred to as "dialects"), cultural and social classes and groups ("sociolects"), other demographics like race, age and gender; and finally, no speaker expresses themselves exactly like another ("idiolects"). 
The variation we describe here may well correlate with such dimensions, because political alignment correlates with many of these dimensions, such as geography ("red states" and "blue states"). More than anything, our results are complementary, not competing with analyses focusing on other dimensions. If geography or age describes a portion of variance in, for example, differences in usage frequencies (Figure \ref{fig_freq} below), then that rather helps piece together puzzles of linguistic variation.

While we base our inferences on public social media data, there are of course other media channels which can and have been studied. 
For example, \textcite{muise_quantifying_2022} argue that US television audiences are much more partisan-segregated than social media users, despite shrinking TV news audiences.
Not all social media behavior is public or accessible either. The advantage of Twitter, compared to some other popular platforms at the time of data collection, was that the public-facing behavior of users (tweets but not private messages) could be easily observed and collected.
However, \textcite{lobera_private_2022} show that platform or communication channel choice can also differ along partisan lines, showing how right-wing supporters in Spanish 2015 elections were more likely to use direct private messaging services for political persuasion activities than the left, who used both public social media and private channels.
The approach we describe can be readily adapted to other social media platforms which facilitate data collection and where users post messages and "follow" or otherwise interact with other accounts.

Furthermore, \textcite{mukerjee_political_2022} caution against overestimating the political nature of social media, arguing that "ordinary Americans are significantly more likely to follow nonpolitical opinion leaders on Twitter than political opinion leaders". However, here we focus on corpora of Twitter tweets posted by two groups of users (see Methods) who either follow left-leaning news outlets and not right-leaning ones (likely "left-leaning users") or right leaning news outlets and not left ones (likely "right-leaning users"). We consider this a proxy for political preference.

It has been argued that using "purely correlational evidence from large observational [social media] datasets" is risky and prone to spurious results \parencite{burton_reconsidering_2021}. Indeed, complimenting "big data" evidence with other approaches has proven fruitful \parencite{kaiser_partisan_2022}. In line with this view, we complement machine learning driven findings with a smaller scale annotation exercise probing the perceived meaning of a subset of words and emoji using human annotation. %

Our contribution is both methodological and exploratory. 
We build on previous research and operationalize the bias of large news media outlets to delineate right-leaning and left-leaning subcorpora of a large sample of tweets. 
We exemplify how a combination of unsupervised, mostly language-agnostic statistical and machine learning driven methods (including generative large language models or LLMs), enhanced by systematic data annotation, can be used to make sense of large quantities of textual social media data to estimate linguistic divergence within a community under conditions of polarization.  
The product of applying these methods is a mapping of lexical and semantic similarities and differences between the "left" and "right" in the United states --- in terms of topics of conversation, usage frequencies of words and emoji, estimated sentiment, and the potentially diverging meaning of everyday words. This allows us to estimate an answer to the question of how much English in the US has diverged across the left-right divide. We find that there is notable divergence in topics and themes of conversation, but also to some extent in lexical semantics.

\section{Methods and Materials}

Our dataset is a corpus of 1,483,385 short posts (or "tweets") and roughly 20 million words on the social media platform Twitter, posted by 10,986 unique users from the United States, between February and September 2021. In the sections below, we describe how these users were selected (\ref{sec_userselection}), with particular attention to the media bias categories which determined whether tweets were categorized as "right-leaning" or "left-leaning" (\ref{sec_accounts}). 
Before presenting our analysis of the final dataset, we describe criteria for excluding individual tweets and pre-processing of the corpus to exclude some aspects of the data (e.g., hashtag symbols, links, audiovisual data; see \ref{sec_cleaning}). 

\subsection{Sampling users on Twitter}\label{sec_userselection}

Users were selected using the following criteria:
\begin{enumerate}
    \item User must follow accounts in one media outlet category to the exclusion of accounts in the other category (detailed in Categorizing media outlets, below)
    \item User must self-identify as being in the US, as indicated by the Twitter API. Users who did not mention a location and have geolocation settings disabled were excluded.
    \item User must be reasonably active, operationalized as: their account being created no later than February 2021, and having tweeted at least 10 times during the observation period.
    \item User must have some engagement with other users: following at least 10 accounts, being followed by at least 5 accounts, and their tweets having a likes to tweets ratio above a threshold of 0.03.
\end{enumerate}

Using a ratio in the final step rather than a raw count allowed us to include users across the spectrum of popularity and volume of activity - users included in the dataset may have had as little as ten tweets and three likes during the observation period, but this also ranged into the thousands. While we placed no upper limit on the like to tweet ratio, tweets within each user profile were ranked by engagement (sum of likes and retweets; in the case of ties, preferring longer tweets) and only the 700 highest ranked tweets by any individual user were included. This ensured our sample was not dominated by individual super users. Overall, this resulted in a total of 11,071 users in the US. Below, we turn in more detail to the first constraint outlined above, before detailing further text cleaning procedures which removed a further 85 users from the sample, resulting in a final sample of 10,986 users.

\subsection{Categorizing media outlets}\label{sec_accounts}

Previous research using social media data to examine political bias has used various strategies to assign a political category to users. Some research uses self-identification, for example by focusing on prominent individuals or smaller samples of prolific public figures with already known political affiliation \parencite{wignell_twittering_2020, chin_evaluating_2022,xiao_detecting_2022,chin_evaluating_2022}, or collecting data from defined subsections of platforms or discussion forums as the niches or samples of interest \parencite{altmann_niche_2011,stewart_making_2018,soliman_characterization_2019}. Other approaches rely on user characteristics or behaviour, using geographical region where geo-location is available \parencite{louf_american_2023}, sampling data by topically relevant keywords or hashtags \parencite{demszky_analyzing_2019,chen_polarization_2021,oakey_phraseology_2022},
categorizing user-generated content \parencite{fraxanet_unpacking_2023}. Here, we usee the general strategy of delineating users based on what kinds of other account they follow on social media  \parencite{an_visualizing_2012,sylwester_twitter_2015,wang_how_2017}. 
We extend this approach in the following way (elements specific to our study in brackets):
\begin{enumerate}
    \item Use a defined set of (US) news media organizations, categorized by political bias (AllSides); 
    \item Find their accounts on the platform of interest (Twitter); 
    \item Mine their full lists of followers;
    \item Group these follower users according to which accounts they do but also \emph{do not} follow;
    \item Mine the posts (tweets) of these users, yielding a subcorpus of text for each group.
\end{enumerate}

\begin{figure}[htb]
	\noindent
	\includegraphics[width=\columnwidth]{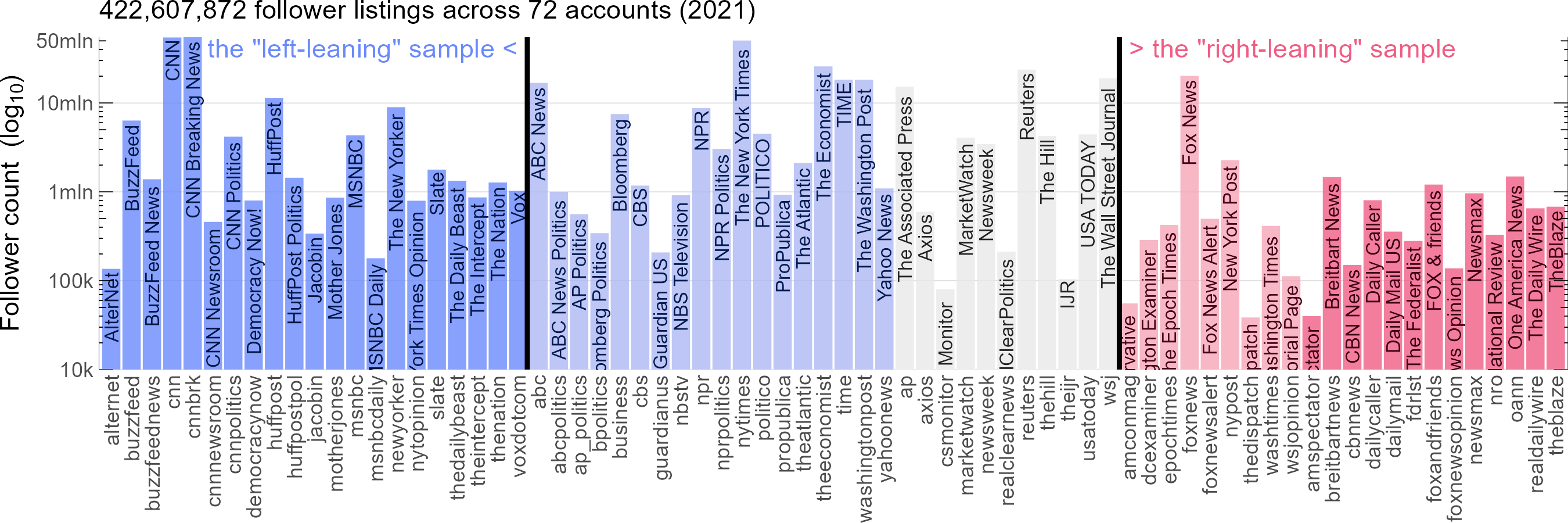}
	\caption{The follower counts of the 72 news accounts on Twitter, grouped and arranged according to their corresponding AllSides (2021) media bias ranking, as left, lean left, center, lean right, and right (alphabetically within each group). The account username is displayed on the axis, the full display name on the bars.
	}\label{fig_accounts}
\end{figure}

We use the AllSides media bias rankings \parencite[][]{allsides_allsides_2021} as a basis to categorize news sources in terms of their political bias (version 4, current at the start of the data collection in 2021; see Figure \ref{fig_accounts}. AllSides media bias rankings are based predominantly on multipartisan editorial review of media outlets combined with an annual, large-scale bias survey of thousands of people in the US from across the political spectrum \parencite{allsides_allsides_2022}. We focus here on the subset of prominent media outlets featured in AllSides' yearly "Media Bias Chart", which categorizes outlets into "left", "lean left", "center", "lean right" and "right". We identified 72 Twitter accounts representing these outlets, listed in the Appendix (note that some outlets have more than one account). 

We use these accounts to categorize users in the following way. First, we assume that following an account is an indication of preference for a news source, as following (essentially subscribing to) somebody, on a live feed-centric platform like Twitter, makes it considerably more likely to be exposed to their content on the platform. In itself, this is unlikely to be a good proxy for political preference: for example, many left-leaning users may follow left-leaning outlets \textit{and} right-leaning outlets, in order to see ongoing discourse on the "other side". However, the premise of our categorization includes an additional criteria: a user who follows left-leaning outlets \textit{and only} left-leaning outlets is likely to be tweeting within left-leaning circles on the platform (likewise, a user who follows right-leaning outlets to the exclusion of left-leaning ones is likely to be tweeting within right-leaning circles). In short, a user following certain news sources with bias A, but not others with bias B, is taken as proxy indicating the user's activity sits more in sphere A than sphere B.

We define the "left" aligned group (colored blue in the graphs) as users who follow at least two accounts in the AllSides "left" category, but do not follow any accounts in any other category. We define the "right" aligned group (colored red in the graphs) as users who follow at least two accounts across the "lean right" and "right" categories, but do not follow accounts in any other category. The colour choices here are aligned with general conventions widely used in reporting and visualization about US politics. Note that this may seem unintuitive particularly to readers familiar with other political systems (e.g., particularly UK political contexts, where Labour [left] is generally red, and the Conservatives [right] are generally blue).

The reason for this slightly asymmetric grouping -- the inclusion of both "right" and "lean right" outlets, but only "left" outlets -- is illustrated in Figure \ref{fig_accounts}: 
more left-aligned accounts from the ranking are represented on Twitter, with more followers on average. This may be related to findings that Twitter users overall are more left-leaning \parencite{wojcik_sizing_2019,pew_research_center_differences_2020}, despite the fact that Twitter's own research shows that right leaning content is more likely to get promoted algorithmically \parencite{huszar_algorithmic_2022}.
Additionally, the boundary between "lean right" and "right" is perhaps more porous, evident for example from the movement of Fox News from the "lean right" to "right" category in subsequent (2022) iteration of the Media Bias Chart. Note that we only consider larger outlets categorized by AllSides: a user may follow smaller news accounts not considered in our categorization process. 

This approach allows us to contrast two subcorpora of tweets with fairly clear and opposing preferences in news sources, and excludes people who consume a balanced news diet or atypical users such as journalists who may follow accounts across the spectrum for professional purposes. 
One downside of this approach is that it requires mining entire follower lists to be able to execute the set operations described above (the does-not-follow part in particular) --- which can be time-consuming, depending on their size and data access speeds of a given platform or API. Then again, this can be entirely automated. Some lists in our sample are quite large, e.g. CNN had 54 million followers at the time of data collection. An upside of the approach is that it allows for starting from users (and then mining their posts and data), instead of requiring the entire corpus of content to be acquired or mined beforehand \parencite[cf.][]{fraxanet_unpacking_2023}. Overall, our implementation churned through the follower lists of the 72 media accounts (totaling 422,607,872 user listings) for about a month between June and July 2021. Using the user-based constraints described here and above, in addition to tweet-based constraints described in more detail below, this yielded two roughly equal subcorpora of 750,180 tweets by 6201 left-leaning users and 733,205 tweets by 4785 right-leaning users.

While tailored here for the Twitter platform and its limitations and affordances, Twitter/X recently restricted access to its research API in ways that will have consequences throughout social science \parencite{ledford_researchers_2023}, including placing limits on the direct replicability of the current study. However, we emphasize that this general approach outlined here is in principle applicable to any kind of (social) media data where the following can be identified: 

\begin{enumerate}[label=(\Alph*)]
    \item An entity or group of entities with an identifiable polarity or bias of interest, and a large enough following or subscriber base (e.g. news sources, popular social media accounts, platforms, forums, etc.) 
    \item The audience, as identifiable users or subscribers
    \item Identifiable links between (A) and (B) in the form of following status, subscription, membership, frequent interaction, etc.
\end{enumerate}

\subsection{Tweet selection and text filtering}\label{sec_cleaning}

The profiles of users meeting the criteria described above were mined for tweets written by the user between February and early September 2021 (including tweets, quote tweets and replies). First, tweets which were not written in English according to the Twitter API were automatically excluded. In addition, irrelevant parts of tweets were modified, or irrelevant tweets were excluded from the dataset based on the following:

\begin{enumerate}
    \item Formulaic uninformative elements of tweets (e.g., AM-PM times of the day, URLs, and tagged usernames indicated by the @-symbol) were removed
    \item Punctuation was removed from tweets (except punctuation-based emoticons).
    \item Hashtags were treated as normal text, i.e., leading \# removed.
    \item Sequences of whitespace greater than a single character were replaced with a single space, and variation in the use of case was removed; all text was converted to lower case.
    \item Variable-length internally reduplicative expressions (e.g. \textit{hahaha}, \textit{hmmm}) were set to uniform lengths. 
    \item Audiovisual information (e.g., images, videos) was removed from all tweets.
    \item Modifier symbols (gender, hair and skin tone) were stripped from emoji.
    \item Tweets containing keywords associated with automated content (e.g. people activating automated services like the "ThreadReaderApp" or "RemindMeOfThis" bots via tweet) were excluded.
\end{enumerate}

Each of these steps was a deliberate choice to make the data feasible to use, and we briefly justify some of these choices here. First, URLs, tagged user names, and times do not reliably contain lexical or semantic information and were thus removed as they were unrelated to our aims in analysis. As we are primarily interested in the lexicon and not syntax, punctuation is removed from the processed tweets (except punctuation-based emoticons). We removed the hashtag \# symbol, but retained the text of the tags in place. While hashtags sometimes follow the body of a tweet, in other cases they are used to tag words within usual sentence structure - we retained the text of hashtags in order to retain sentence structure where this is the case, and we assume the meaning of a word with or without a hashtag to be roughly the same. Given the moderate size of our corpus, we chose not to consider variation in case, focusing instead on lexical and semantic variation. Making variable length expressions like \textit{hahaha} and \textit{hmmm} uniform in length allowed us to consider their use across groups more effectively. Finally, including all aesthetic variations of emoji would greatly increase the complexity of comparisons; in line with our overall aim to detect general semantic patterns, this variation was removed.
Finally, a handful of accounts with anomalous tweets were removed from the dataset (i.e. those repeatedly posting identical or promotional tweets; 85 users and all their total of 17,692 tweets, including the anomalous and all other tweets). 

Prior to all these filtering steps, the corpus had 21,327,634 million whitespace-separated tokens with a type-to-token ratio (TTR) of 0.05, meaning that for every hundred tokens there were approximately five distinct word types, which is very high. For comparison: the 2016-2017 segment of the written part of the Corpus of Contemporary American English \parencite{davies_corpus_2008} is 18.6M words; TTR for its lemmatized version is 0.008 and unlemmatized 0.01.
After filtering, cleaning and lemmatizing, our final corpus came down to 20,357,194 tokens with a TTR of 0.01, consisting of 1,483,385 tweets from 10,986 users.

\subsection{Lemmatization}

While most people might think the question of what it means to be a word is a trivial one, linguists disagree substantially on what counts as a word or term for comparative purposes, and on how this should be operationalized in different contexts \parencite{dixon_word_2003,haspelmath_indeterminacy_2011}. Nonetheless, this is often not given much attention in computational lexical change literature, which often relies on more or less white space-based tokenization \parencite[][]{hamilton_diachronic_2016,feltgen_frequency_2017,schlechtweg_semeval-2020_2020}. However, using simply tokenized raw text risks losing key lexical and semantic relatedness between similar strings, for example, that both both \textit{runs} and \textit{running} are uncontroversially instances of the verb \textit{to run}.

Lemmatization is the process of stripping strings of morphological inflection and collapsing them in terms of their root form, in order to detect string tokens which might share a root lexical form. For example, both \textit{runs} and \textit{running} are instances of the root \textit{run}. This process is often used for lexical and semantic analyses, as it allows the detection of similarity between e.g., \textit{runs} and \textit{running}, that would otherwise be lost with pure white space tokenization. In particular, this process allows us to make more accurate frequency estimates of root lemmas (by e.g., summing the frequency of \textit{runs} and \textit{running} alongside \textit{ran},\textit{run} etc). 

Overall, we use the term "word" to refer to various meaningful units: words in the dictionary sense, proper nouns, hashtags, emoji, emoticons, and the concatenated collocations. However, lemmatization suits our main goal of ultimately comparing semantic concepts (such as the activity of running, regardless of whether it is expressed as a noun or a verb), rather than morpho-syntax, particularly for our topic, word frequency and semantic divergence analyses (for sentiment analysis, the text was not lemmatized). Here, we use the English-specific tools in the Python spacy library \parencite{honnibal_spacy_2017} for tokenization (separation of strings, e.g. by white space) and lemmatization.

\subsection{Word embeddings}

First, we use word embeddings to estimate semantic divergence across the entire lexicon represented in our corpus. 'Semantic divergence' quantifies the extent to which a single lexical item is used in different ways;
between two or more communities (as represented by corpora).
High semantic divergence means a given word is used in different senses in the different groups or communities.
Specifically, we aim to explore whether semantic divergence occurs between right-leaning and left-leaning tweets within our corpus.

Following previous research, we use a type-based model which assigns a fixed vector to each word \parencite[fastText, essentially word2vec with subword information;][]{bojanowski_enriching_2017}. This consists of training two separate embeddings on the left-leaning and right-leaning subcorpora, then normalizing and aligning the vectors \parencite[using the Orthogonal Procrustes approach; cf.][]{hamilton_diachronic_2016,schlechtweg_wind_2019}. 
Divergence is estimated via pairwise cosine similarity in the aligned embedding: high similarity across aligned embeddings indicates low semantic divergence, while low similarity indicates high divergence.
This approach performs well in detecting diachronic lexical change \parencite{schlechtweg_semeval-2020_2020} which is analogous to our case of detecting synchronic divergence. Type-based embeddings are easy to implement and interpret, yet have been shown to outperform more recent resource-intensive models in these kinds of tasks \parencite[e.g., BERT-like token-based approaches driven by pretrained LLMs; but cf.][]{kutuzov_contextualized_2022,rosin_temporal_2022}.

For both word embeddings and frequency comparisons, we exclude words with infrequent usage in the comparison: a word must occur at least 100 times in both the left-leaning and right-leaning subcorpora to ensure reasonably reliable semantic inference. This leaves 3582 words (lemmas) and emoji. 
We optimize the training parameters by maximizing the average of self-similarity of words between the two embeddings (after the alignment step). The assumption is that since this is still the same language, most word pairs should have similar vectors. The final models have dimensionality of 50, window size 5, minimal frequency of 5, and 5 training epochs (training for too long easily leads to overfitting and weakly aligned embeddings, likely due to the moderate size of the dataset).

\subsection{Semantic annotation by humans and machines}

We use a human annotation to evaluate the perceived semantic divergence of a subset of words and emoji detected by the model as being particularly divergent. 
Unsupervised machine learning approaches, such as the model described above, are difficult to evaluate in terms of their accuracy.  In the case of word embeddings, the model results may reflect genuine semantic (dis)similarity, and/or rather variation in context. Compared subcorpora may also diverge considerably in discussed topics.
While training our models from scratch sidesteps the issue of possible biases of large pre-trained models, they may be susceptible to frequency biases \parencite{wendlandt_factors_2018} and sensitive to parameterization. Tests on our data with different training parameters, for example, yielded slightly different results in terms of most divergent words.
We therefore select a subset of words and emoji for model validation, using both human and LLM-driven annotation.

This takes the form of a semantic annotation exercise adapted from the DURel framework \parencite{blank_prinzipien_1997,schlechtweg_diachronic_2018}. The advantage of this annotation framework, originally demonstrated on diachronic data tasks \parencite{schlechtweg_diachronic_2018} but equally applicable here, is that it does not require the annotators to be speakers of the specific variety, just proficient speakers of the language the variety comes from or is closely related to.
Annotators are presented with pairs of sentences or passages where the target word of interest occurs. The task is to rate the similarity of the target on a scale of 1 (unrelated) to 4 (identical meaning) given the context. 
Manipulating the subcorpus from which each sentence in a pair is drawn allows for the estimation of both (dis)similarity or divergence (scores of example sentences from different subcorpora) and in-group polysemy (scores of examples from the same subcorpus). This is informative, as the combined results indicate if a given word usage differs on average between subcorpora just because it is polysemous and its different senses are just used with different frequencies --- or, if a given target word refers (only) to different, unrelated concepts (due to semantic divergence across groups represented by the subcorpora; more akin to homonymy). In our exercise, both co-authors independently provided DURel scores for the test set of passages (partial tweets to speed up annotation; a context window of up to $\pm60$ characters around the target). We evaluated 8 target words, 40 unique passages each, which were (randomly) combined to produce 20 left-right pairs, 10 left-left and 10 right-right pairs, for a total of 320 paired comparisons completed in a random order.

When sampling the corpus for examples for this exercise, we only consider tweets with enough context ($\geq70$ characters and $\geq10$ words in length, TTR $\geq0.6$) and exclude those with irregular use patterns (ratio of the sum of 2 most frequent letters to total length $<0.4$; ratio of Capitalized words to uncapitalized $<0.5$). Target nouns are allowed to be in plural form, but not surrounded by hyphens, as these could be meaning-altering compounds. Tweets were randomly sampled from the remaining corpus, including a maximum of one tweet per user, preferring longer tweets to ensure roughly uniform stimuli lengths.

In addition to this, we had a generative LLM complete the same task, exploring the feasibility of using current-generation LLMs to estimate divergence and act as data annotators \parencite[following][]{ziems_can_2023,huang_is_2023,gilardi_chatgpt_2023}. We use OpenAI's gpt-4-0613 model via its API \parencite{openai_gpt-4_2023}. This model is also referred to as "GPT-4", which also powers the popular ChatGPT chatbot. We used the following prompt: "The target words in <x> tags in sentences A and B are spelled the same, but their meaning in context may be similar or unrelated (homonymy counts as unrelated, like bat the animal and bat in baseball). Rate meaning similarity, considering if they refer to the same object/concept. Ignore any etymological and metaphorical connections! Ignore case! Ignore number (cat/Cats = identical meaning). Output rating as: 1 = unrelated; 2 = distantly related; 3 = closely related; 4 = identical meaning. [followed by the two example passages]".

\FloatBarrier
\section{Results}

\begin{figure}[htb]
	\noindent
	\includegraphics[width=\columnwidth]{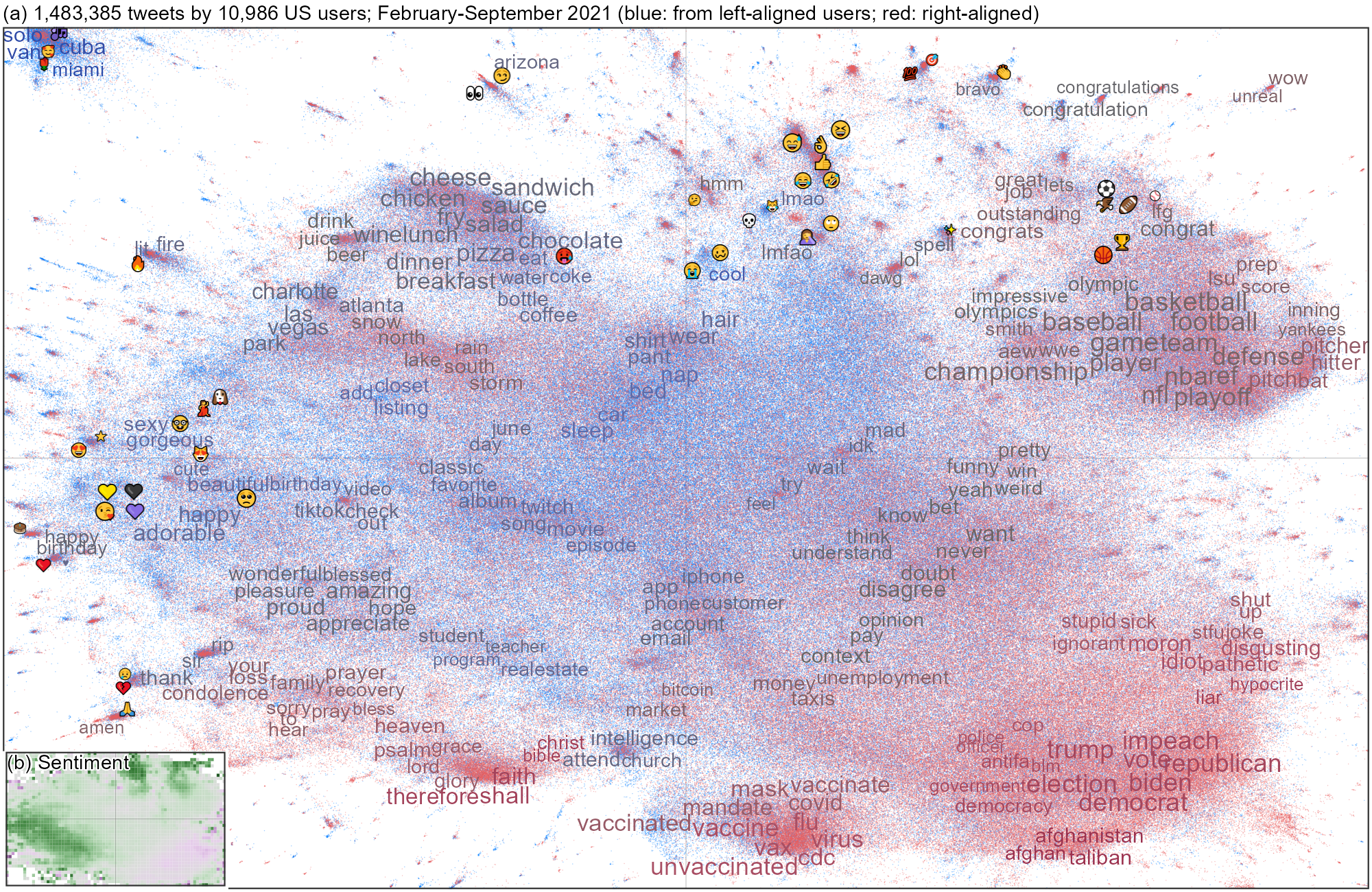}
	\caption{(a) 1.5 million tweets authored in the US in 2021, colored by estimated political alignment (blue is left-leaning, red is right-leaning). Tweets close together are semantically similar.
    Topical keywords have been plotted over dense clusters (colored similarly, by the share of red vs blue user tweets in the cluster). 
    Some topics like food and birthdays are discussed regardless of political alignment. 
    The blue areas stand out with everyday life topics (keywords like \textit{sleep, car, birthday}).
    The top left blue corner are mostly bilingual tweets containing Spanish. 
    Some political figures, religion and vaccination related topics appear more popular in the right-leaning subcorpus. 
    The inset (b) is a heatmap of the same UMAP, colored by the average estimated sentiment of the tweets (purple negative through gray neutral to green positive; see the sentiment analysis section for details). 
    The political tweet cluster in the bottom right again stands out as notably more negative.
    This map illustrates how groups of people of opposing political alignment in the US, while sharing some topics of conversation, noticeably diverge in others.
	}\label{fig_topics}
\end{figure}

Figure \ref{fig_topics} depicts our corpus of tweets, colored by the estimated political alignment, 
arranged by semantic or topical similarity, 
as a UMAP dimension reduction \parencite{mcinnes_umap_2018} 
of a doc2vec (or paragraph2vec) text embedding \parencite{le_distributed_2014}. 
UMAP provides a two-dimensional topography of the full 50-dimensional topic model. The doc2vec model uses fasttext embeddings \parencite{bojanowski_enriching_2017} as input, here trained with the same parameters as the semantic models discussed in the Lexical-Semantic Divergences section below.
The sporadic words and emoji on the graph are salient keywords 
(frequent in these topics, calculated via term frequency-inverse document frequency or TF-IDF scores) 
of local DBSCAN tweet clusters \parencite[the top2vec approach; cf.][]{angelov_top2vec_2020}.
This allows for a first impressionistic birds-eye view of the entire corpus and the topical clusters within it.

While some areas of the topical map contain tweets from both sides (mix of blue and red dots), some predominantly red and blue areas are immediately visible. This indicates that the distribution of conversation topics is not entirely independent of political leaning. 
One way to quickly test this impression is to train and test a classifier to predict the (estimated) alignment of the author of each of the 1.5 million tweets. 
A prediction accuracy above chance would indicate a discriminable difference between the left and right-leaning subcorpora. 
We use a simple model, Linear Discriminant Analysis, with the 50 latent dimensions of the doc2vec model as the predictor variables. It is able to predict the previously estimated alignment of left or right (see Methods) with an accuracy of 64\% (or 27\% kappa score, on the roughly 50-50 class split; bootstrapped accuracy estimate). While this accuracy is far from perfect, it sits well above random chance, meaning that there is enough topical or usage divergence across users in each subcorpus to guess their news diet preferences (and by proxy, political preferences) with reasonable accuracy. 
It also mirrors previous research comparing tweet content (both text and images) of followers of Donald Trump and Hillary Clinton and reporting a similar classification accuracy \parencite{wang_how_2017}.
In the following sections, we investigate this in further detail by looking at usage frequencies, estimated sentiment and lexico-semantic divergence.

\subsection{Usage frequency differences}

Word frequencies in comparable corpora, differentiated by e.g. time period, genre or social group, can provide insight into the average usage patterns of the speakers whose utterances make up the data, including social media data, as shown in previous research \parencite{grieve_mapping_2018,louf_american_2023}.
We employ the following operationalization to provide a straightforwardly interpretable overview of aggregated usage differences between our left-leaning and right-leaning subcorpora. 
To focus on words with reasonably reliable frequency estimates and to reduce possible effects of idiosyncratic usage, we simply filter the lexicon here to only include words which occur $\geq200$ times in either subcorpus, $\geq300$ times in total, used by $\geq200$ users in total, with a users to token frequency ratio $\geq0.05$.

\begin{figure}[htb]
	\noindent
	\includegraphics[width=\columnwidth]{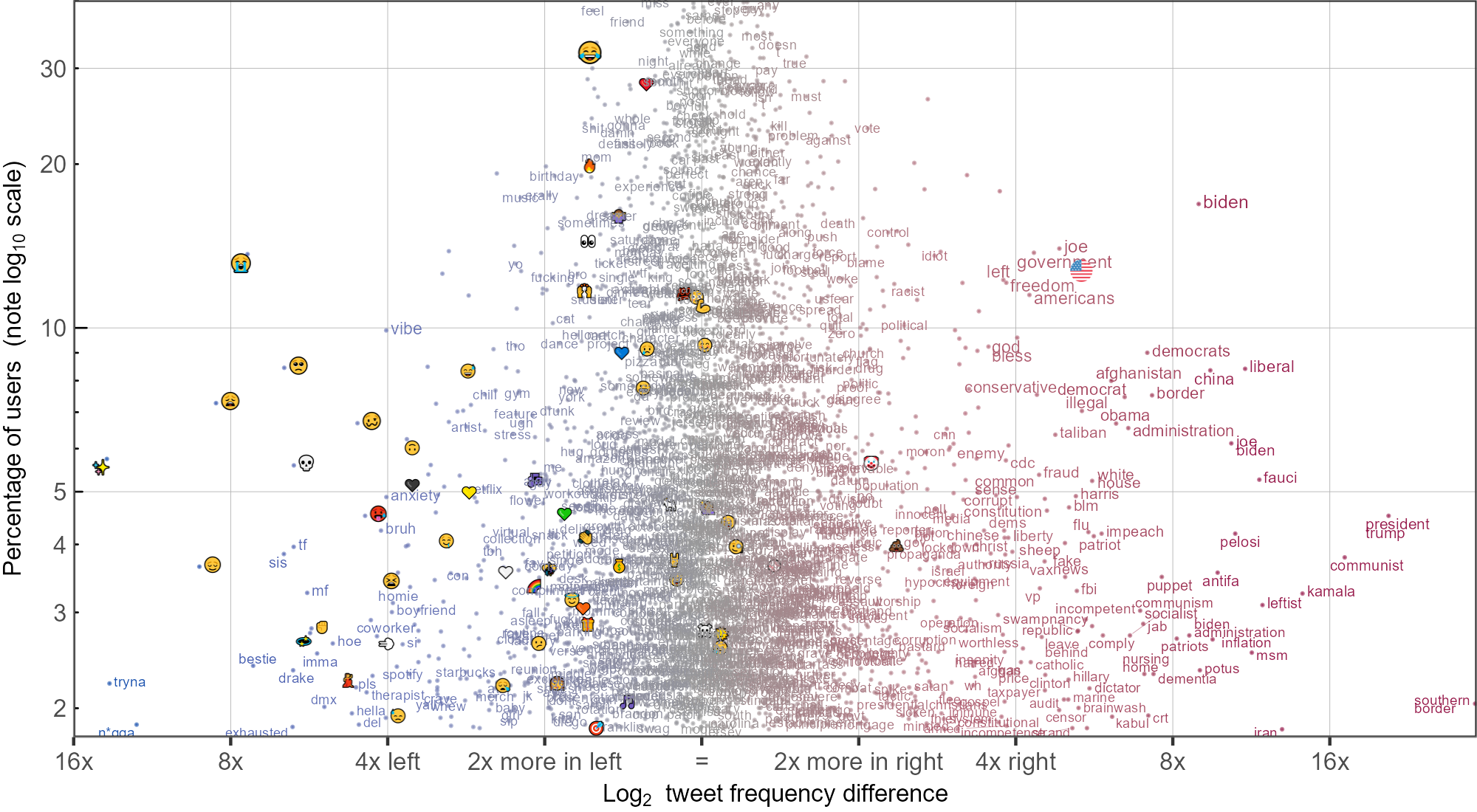}
	\caption{Word usage frequency differences between the left and the right, February-September 2021. The difference is on a normalized $\log_2$ fold scale, straightforwardly interpretable as multiplicative difference. The vertical axis reflects the overall frequency of a word, as percentage of users whose tweets contain it (clipped at 30\%, as more frequent words like function words don't display large differences --- the two groups are still speakers of the same broad variety of English). The left-leaning corpus stands out with more non-standard e.g. \textit{sis, bestie, bruh, wanna} and more emoji ("sparkles", various faces) --- with the exception of the "clown", "poo" and the "US flag" emoji. Political terms and names such as \textit{Biden, Democrats, liberal} etc. are more frequent in the right-aligned corpus.
	}\label{fig_freq}
\end{figure}

For the comparison itself in Figure \ref{fig_freq}, we use the number of tweets a word occurs in as the frequency, normalized by the number of tweets in the respective subcorpus. Tweet frequency instead of token frequency allows for the meaningful comparison of conventional words and emoji on the same scale --- as the latter have reduplicative usage properties, unlike most words. For example, the laughing-with-tears emoji (top middle in Figure \ref{fig_freq}) occurs in multiples, in about 42\% of the tweets where it is present, whereas the median is 4\% among short words (2-3 characters) and 3\% among longer words.

The frequency difference metric in Figure \ref{fig_freq} is on the logarithmic scale (being more informative than linear given the Zipfian nature of word frequencies), 
as $log_2(f_{w_r} / f_{w_l})$, the logarithmic difference for each word, between the frequencies in the respective left- and right-leaning subcorpus.
The binary logarithm value has the convenient advantage of still being interpretable as fold or multiplicative difference for integer values, e.g. the score of a word that is used in 200 tweets per million tweets in the right-leaning subcorpus, and 100 on the left, $log_2(200 / 100){=}1$, is twice as frequent, $log_2(400 / 100){=}2\ $ is 4x more, etc.

The words with the most different frequency distributions between the subcorpora tend to be political figures and politically charged terms for the right, and emoji for the left. Across all spellings i.e. lower and upper case, \textit{Joe Biden} is used about 10 times more on the right (used by 674 users; as just \textit{Biden} 9x, 1856 users out of the total of 10,986 users in the corpus). 
In general, as a reminder, we lowercased and lemmatized the corpus, so all frequencies discussed here refer to the sum of occurrences of a term that may or may not include various spellings and morphological variants such as singulars and plurals.

Here and in the following, we will present some illustrative example data from our tweet corpus. To be on the safe side, these are however synthetic, either composite or rephrased examples, as publishing original tweets verbatim would make the users and therefore their inferred political leanings identifiable, which may be problematic.

Despite being blocked from the platform in January 2021 following the events of January 6th, \textit{President Trump} still appears in 1659 tweets in our corpus. The term appears almost 21 times more frequently in right-leaning tweets, but is nonetheless only used by a vocal minority of 496 users ($<5\%$ of our sample; 460 of them right-leaning)
Other names and terms more frequent among the right include 
\textit{communist} (17.4x more on the right, 416 users total, 372 right-leaning),
\textit{Fauci} (11.7x more on the right, 578 users in total, 514 right), 
\textit{liberal} (11x, 924, 778 right), 
\textit{border} (7.3x, 826, 652 right),
\textit{America} (3.2x, 2097 users, 1347 right-leaning).
Again, it is clear that some frequencies may be driven by vocal users rather than large differences in users who would use a given word in general.

Some (synthetic) examples include:\\
"The Democrats are not liberals, they are fascist totalitarian communists, there is nothing liberal about them", \\
"Beijing Biden is the one causing the Border Crisis. He is opening Our Borders to traffickers and killers", \\
"Happy birthday America! The beacon of freedom! \#4thOfJuly \#GodBless [US flag emoji]" 

In the left-leaning subcorpus, we find several emoji which are more frequent than in the right-leaning subcorpus: the "sparkles" (13.8x more on the left; 631 users in total, 536-left-leaning), "crying face" (7.4x, 1501, 1154 left),
and the "skull" emoji (6x, 614 total, 475 left).
In addition, we find several terms used more frequently in the left-leaning subcorpus:
\textit{vibe} (4x, 1087,  849 left), 
\textit{wanna} (3x, 1520, 505 on the left),
and additional vernacular usage by smaller groups such as
\textit{sis} (6.3x, 422 users total), 
\textit{tf} and \textit{af}  (largely shorthands for \textit{the f*ck} and \textit{as f*ck}, 6x and 3.6x on the left, 435 and 686 users in total, respectively). Some synthetic examples: \\
"The best things in this life are not things. [sparkles] Grateful to you all for the smiles. \#fridaymood", \\
"ppl really just be on their couch, no medical background, just sage and vibes, tryna disprove COVID. get vaxxed please [5 crying face emoji]", \\
"Haha sis don't play me like that".

The only emoji used by more than 5\% of the sample that are noticeably more frequent in the right-leaning corpus, are the "clown face" (2.1x) and the "US flag" emoji (5.3x more). Not all emoji are divergent in usage: the simple "red heart", tweeted by 3073 users, appears only 1.3x more on the left, while the "biceps" muscle emoji and the "face with rolling eyes" (1207 and 1372 users total) occur almost equally on both sides.

\FloatBarrier
\subsection{Sentiment differences}

Sentiment or emotional polarity in a corpus could be either interpolated from a manual analysis of a smaller sample, or inferred from rough estimate of a machine learning or statistical analysis of the entire dataset. 
We opt for the latter, employing the VADER (Valence Aware Dictionary for sEntiment Reasoning) model, due to its robust performance also on social media data \parencite{hutto_vader_2014}. Its compound scores, based on the individual words and estimated valence, range in $[-1,1]$. 
For example, the sentence "This is great!" scores 0.66. Adding a smiley ":)" raises it to +0.81, while "This is not great!" gets -0.51. 

To further illustrate the sentiment model, tweets like 
"Is the gov really paying for this crap? all so FAKE all full of LIES :(", 
and "My vote would never go to some senile old man! [4 screaming face emoji] Sorry for you though!!! [2 crying emoji]" 
both get strongly negative scores below -0.9 (these and the following are synthetic examples, as above).

This type of sentiment model, mapping text on an abstract negative-neutral-positive scale, has the downside of marking texts with potentially very different meaning and intention with a similar sentiment score, if the content includes words listed with a similar polarity in the model (while VADER does take negation into account, like all NLP models, it can misinterpret human sarcasm and irony). 
The upside is that its results are fairly straightforward to interpret, if its limitations are kept in mind \parencite[including the nature of the data the sentiment or stance is inferred from, cf.][]{joseph_misalignment_2021}.

The sentiment inset (b) in the overview Figure \ref{fig_topics} is based on the application of the VADER model. These results suggest that there might be differences in that aspect between the groups, as predominantly right-leaning areas of the topic map are also some of the more negative areas on the sentiment map.
VADER scores both out-of-vocabulary words and those with neutral sentiment as zero, so here we exclude tweets consisting of solely zero-value word-scores (31\% of the corpus; compound scores just averaging at zero are not excluded) for a more precise comparison. The results can be interpreted as differences between the groups in terms of tweets with a detectable polarity.

\begin{figure}[htb]
	\noindent
	\includegraphics[width=\columnwidth]{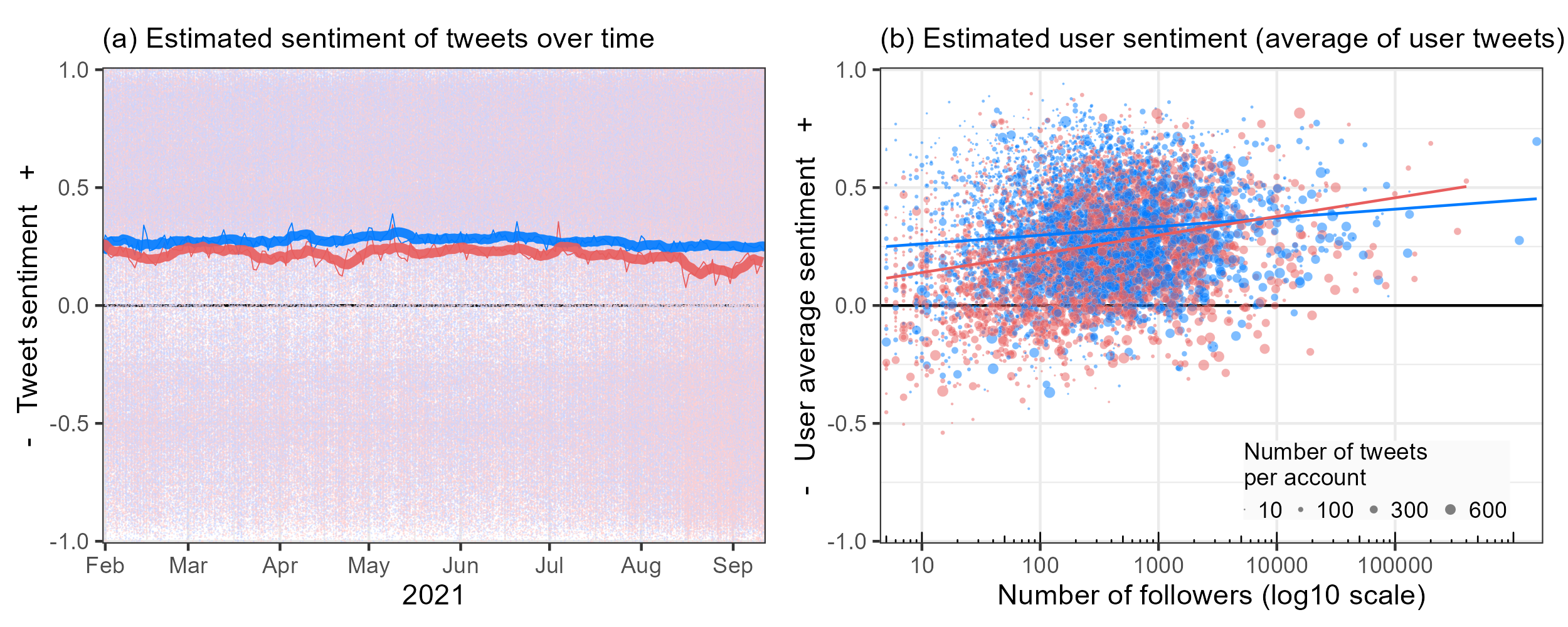}
	\caption{Estimated tweet (a) and user average sentiment (b), negative to positive. Each dot on (a) is a tweet, colored blue for left and red for right (a small amount of vertical noise is added, as many short tweets would overlap due the averaging of word sentiment), superimposed by daily (thin lines) and weekly averages (thicker lines). 
    Panel (b) depicts all the users in the sample, arranged on the y-axis by the average sentiment of their tweets (excluding neutral-only tweets; see text). Right-aligned users are slightly more negative on average. There is a small positive correlation between popularity (number of followers) and average sentiment, more pronounced among right-wing users. User dots are sized by the number of their tweets in our sample. 
    This figure illustrates the two political sides are rather similar in their average social media sentiment, with a slight skew towards the negative among some smaller right-wing accounts.
	}\label{fig_sent}
\end{figure}

The estimates of the model for all the remaining tweets in the corpus are arranged over time in Figure \ref{fig_sent}.a, and averaged per user in \ref{fig_sent}.b. On average, both the red and blue US Twitter appear to be fairly stable over the course of 2021, on average staying more positive than negative. 
While tweets by both sides consistently cover the entire sentiment spectrum, the rolling average of tweets by right-aligned users appears to be slightly more negative (the red line staying below the blue one in Figure \ref{fig_sent}.a). Controlling for user variation using a mixed effect linear regression model with a random intercept for user, the right-side tweets are on average $\beta=-0.07$ lower than the left, $p<0.0001$ compared to an intercept-only model (model assumptions are roughly met, although the dependent variable is bounded).

Figure \ref{fig_sent}.b averages tweet sentiment for each user, and displays the size of their following. Besides the right being a bit more negative, as already apparent before, we find a small yet significant positive correlation between account popularity ($\log_{10}$ number of followers) and averaged sentiment (linear regression, $\beta=0.06$, $p<0.0001$, $R^2=0.03$, i.e., positivity increases by about +0.06 with each order of magnitude of follower count). As indicated by the regression lines in Figure \ref{fig_sent}.b, this effect is somewhat more pronounced among right-wing users (positive interaction between side and followers, $p<0.0001$, model $R^2=0.06$), possibly due to the negative less popular accounts dragging its average down. This is only a small correlation, and there is plenty of positive messaging among small accounts, as well as popular accounts with a near-neutral or negative average. 

Among the popular but negative accounts we find the account of a Republican politician
with 19k followers (at the time of data collection in 2021) and an average estimated sentiment of -0.04.
A negative sentiment estimate can stem from very different messages though. For the latter user, it includes language like the following (synthetic examples).\\
"Joe Biden is the President that every extremist, kidnapper, felon, arms dealer and child molester has always been dreaming of", and \\
"Terminating a pregnancy equals murder. They have chosen murder as their call to arms". \\
For comparison, the tweets of another negative-averaging (-0.03) environmental journalist account with a similarly sized follower base includes text such as
"\#Heat wave in Oregon, fatality count hits 106, a mass casualty incident. \#Climate crisis could endanger billions due to \#malaria and other viruses".
While the lexicon-based sentiment may be similar, the content is obviously quite different.
The largest account in our sample appears to belong to a sports coach with 1.6M followers, who is also among the most positive accounts at +0.7, tweeting mostly various congratulations and happy birthday wishes.

\FloatBarrier
\subsection{Lexical-semantic divergence}

The previous sections dealt with frequencies of words, and sentiment as inferred from the frequencies of words with a certain polarity. We are also interested in the semantics of words, and in particular, if there are large enough discrepancies in the intended meaning of some words between the left- and right-leaning subcorpora for this to feasibly cause communicative misunderstanding, and therefore potentially fuel further polarization.

We approach this using a combination of machine learning driven and qualitative annotation methods.
Using a word embedding model, we can easily estimate the semantic difference between the left and right subcorpora for every word in the English lexicon that is represented and sufficiently frequent in our data (see Methods and materials).
This complements previous work (on the English language) which has focused on a limited vocabulary of interest rather than the lexicon at scale \parencite{bhat_covert_2020}, semantic and usage pattern differences between specific people \parencite{li_speaking_2017} or news sources \parencite{spinde_identification_2021}, and comparable diachronic research \parencite{azarbonyad_words_2017,rodman_timely_2020}.

\begin{figure}[htb]
	\noindent
	\includegraphics[width=\columnwidth]{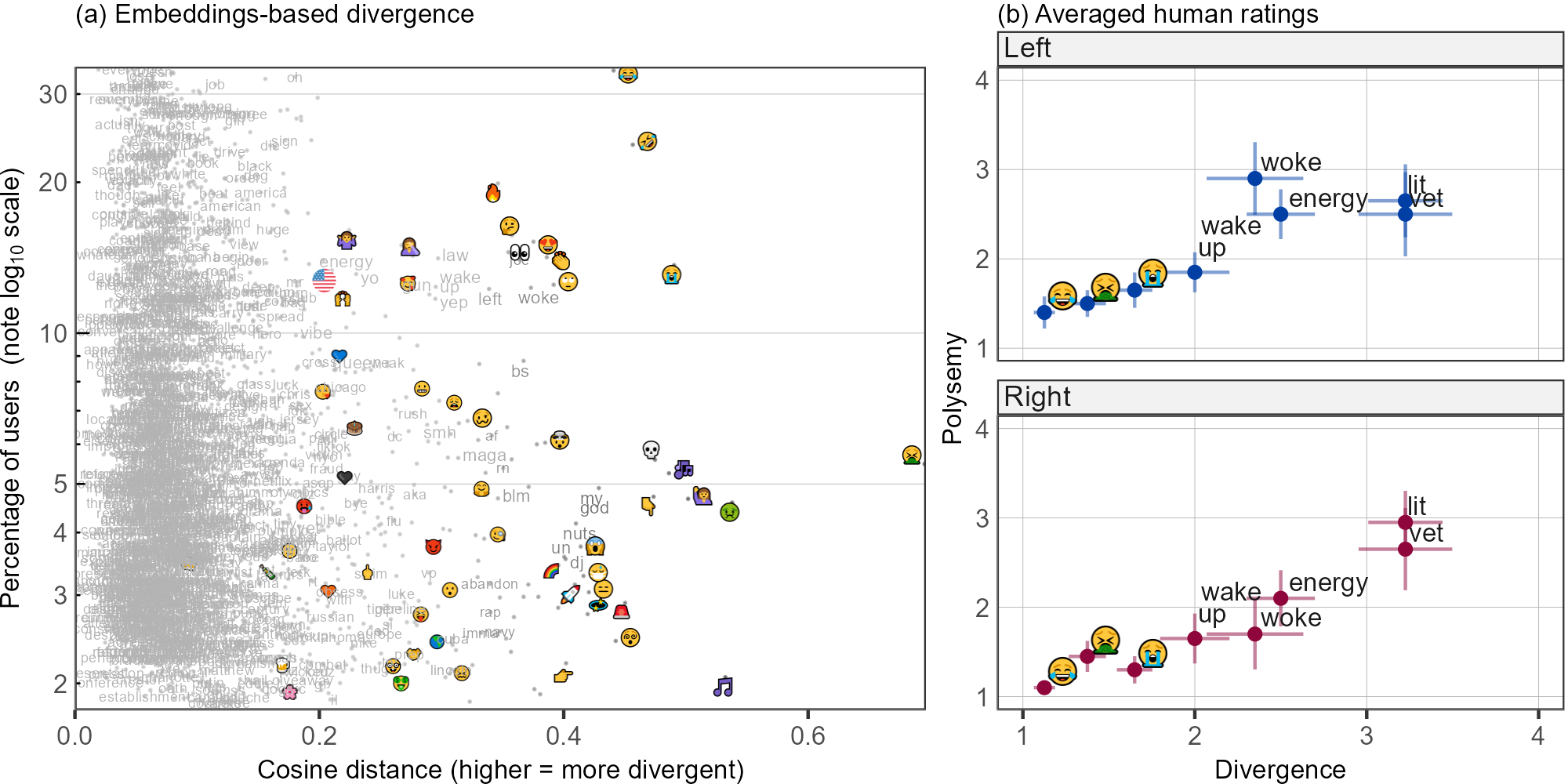}
	\caption{Semantic divergence between the left- and right-leaning subcorpora, quantified via word embeddings (a) and a human annotation exercise on a smaller subset of terms and emoji of interest (b). 
	The word embeddings highlight a number of words that may be either used in differing senses, or at least in highly different contexts. Some of these are used by a small percentage of users (y-axis), while there appears to be divergence also among more frequent terms (e.g. \textit{woke}, various laughing and crying emoji). 
    The annotation results show that emoji are fairly monosemous and used in the same function (therefore likely just differ in context), while words like \textit{lit} and \textit{woke} are indeed used in different senses. 
    The position of the words (averaged divergence scores across annotations; bars show standard errors) on the x-axis of is identical on the two subplots of (b), while the y-axis reflects polysemy within the respective subcorpora --- which is similar, but e.g. \textit{woke} is more polysemous on the left, used to refer to both waking up and being alert to prejudice and discrimination.
	}\label{fig_semantics}
\end{figure}

Figure \ref{fig_semantics}.a depicts the results of applying the Procrustes-aligned fasttext embeddings approach. The vertical axis corresponds to Figure \ref{fig_freq}, the share of users in the sample who use a given word, while the horizontal axis is the semantic divergence, measured as cosine distance (1-similarity) between the vectors of a given word in the aligned embeddings. The most divergent among the more frequent cases are a selection of facial emoji, terms like \textit{woke}, \textit{bs} (largely short for \textit{bullsh*t}), \textit{left}, \textit{lit} (which can refer to lights but also mean "cool, awesome"), and the phrase \textit{wake up}, which can be used literally or figuratively as as a rallying call to pay attention.

The human annotation results Figure \ref{fig_semantics}.b are limited to a test set of 8 targets: the "laughing in tears" emoji, the "vomiting" emoji, the "crying with tears" emoji, the phrase \textit{wake up}, and the words \textit{woke}, \textit{energy}, \textit{lit}, and \textit{vet}. Estimating these results required annotating 320 pairs of example passages; see Methods and materials). The scores depicted in Figure \ref{fig_semantics}.b are a result of averaging the results of the two annotators, who had fairly high inter-rater agreement of $\rho=0.87$ (measured here using Spearman's rho, given the ordinal scale). Both divergence and in-group polysemy are presented as an inverse of the DURel scale, representing how different (dissimilar usages between left and right) and how polysemous (dissimilar usages within left and right) the meaning of each target word is.

Here more divergent words (across subcorpora) are also more polysemous (within their subcorpora), indicating that while the two sides use different senses, they are still mutually intelligible \parencite[this is not surprising given that this is still largely the same language, and also given how meaning extension likely works in diachrony, cf.][]{blank_prinzipien_1997,ramiro_algorithms_2018}. For example, when a right-aligned person uses the word \textit{vet}, they are simply more likely to refer to a (military) veteran, and one on the left to their veterinarian, but both senses still exist on both sides. A complete divergence would be a word located in the bottom right of Figure \ref{fig_semantics}.b --- completely unrelated meanings, and no polysemy that would facilitate sense overlap.

The annotation exercise also serves as a way to partially evaluate the word embedding driven results: $\rho=-0.9, n=8, p=0.002$. The negative correlation, indicating a mismatch, appears to be driven mostly by the emoji, which the embedding approach infers to be moderately divergent, yet human annotators see as fairly similar in usage. Furthermore, the annotation targets were selected from the diverging (right hand) side of Figure \ref{fig_semantics}.a --- the negative correlation is therefore informative about diverging words but not the entire embeddings. This result still highlights that cosine similarity derived from word embeddings captures not only semantic similarity but also contextual or topic differences. Emoji in particular are multi-functional elements that can be used to illustrate, modulate and change the meaning of a text. For example, we observe the "crying" emoji being used to express sadness as well as happy tears; and the "puking" emoji being used literally, to express sarcasm, as a noun, as a verb, and being used in lieu of letters inside a name (presumably to express sentiment towards the person).

As such, unlike many words, emoji can occur in highly variable contexts and functions (without being constrained by syntactic rules like words). Where contexts differ, word embeddings and language models are likely to represent them with differing vectors. Previous research has attempted to infer semantic change in emoji using similar embedding methods \parencite{robertson_emoji_2020}. We would therefore suggest that any such research involving emoji should additionally control for topical variation \parencite[cf.][]{karjus_quantifying_2020}. This does is not to say the current result depicted in Figure \ref{fig_semantics}.a is invalid or uninformative, but it may pick up on signals other than just lexical semantics.

We also experimented with applying a pretrained large language model \parencite[GPT-4;][]{openai_gpt-4_2023} on the annotation task, prompting it with the same DURel annotation instructions to evaluate the semantic similarity of the target word on a 1-4 scale. We find that it achieves moderately good agreement with human annotators ($\rho=0.45$ and $0.6$ respectively). This is lower than the human inter-rater agreement --- partially driven by the emoji, which are indeed difficult to evaluate, as well as to instruct how to evaluate. Nonetheless, without the three emoji, the agreement only rises to 0.54 and 0.66, relative to the 0.87 inter-rater agreement between human annotators. This underscores the limitations of using large language models for complex annotation tasks, and the need to evaluate their output.

Example pairs that require interpreting the conveyed sense of the emoji can look like the following (synthetic examples as above): \\
"Now our sons are off to university. Imma miss the crew [\underline{crying emoji}] but this was the goal all along...to get in and WE DID IT [heart emoji]", \\
"Discovering the hard way why the sauce I ate yesterday is named Red Dragon sauce. Pain pain pain [\underline{crying emoji}]"  (emoji were presented in their original form in the annotation task). \\
Other examples where humans can infer the difference but the LLM can fail are highly contextual, such as this pair: \\ 
"This guy is worried about the notion of white rage, he should really worry about \underline{vet} rage. Soldiers have sacrificed lives and arms and legs for two decades now", \\
"We had a kitten brought in last night and she's struggling today. The follow-up at the \underline{vet} earlier in the morning went fine, but condition deteriorated this evening."

Regardless, these results are promising, especially given the difficulty of this contextually complex yet minimal-context task. Better models and better instructions may well edge the results closer to human performance, as they already have been in some other applications \parencite{gilardi_chatgpt_2023,huang_is_2023,begus_large_2023}. 
Still, the results illustrate the necessity to evaluate machine learning results against human evaluations, but also the potential of enhancing and scaling up the (otherwise highly laborious) human annotation processes using machine learning based tools.

\FloatBarrier
\section{Discussion}

The results on divergence on topics of conversation echo previous research focusing on the differing daily lives of people in the US of opposing political alignment \parencite{brown_measurement_2021,hetherington_prius_2018,rawlings_polarization_2022} --- while naturally many topics overlap, others are segregated, and there are a number of words being used several times more on one side compared to the other. On average, the two subcorpora are similar in tweeting sentiment, although we found a small (yet significant) effect of slightly more negativity on the right leaning subcorpus. The topics where negativity tends to occur tend to be predominantly political topics (see Figure \ref{fig_topics}).

We also probed lexical semantic divergence using two machine learning models and a systematic data annotation approach. This revealed that while there are some words, emoji and phrases with a diverging or at least variable meaning, the cases we tested via manual annotation exhibit differing distributions of sense usage in polysemous or homonymous word forms, rather than divergence in progress. This is not to say that given time, word senses in American English in the US may not diverge enough to begin to cause genuine misunderstanding.

\subsection{Limitations}

Our annotation exercise was limited to eight target words and two annotators (the authors). Provided sufficient resources, the DURel framework we used here lends itself well to be scaled up to a larger, potentially crowd-sourced online experiment \parencite[with care given the issues with such platforms, cf.][]{cuskley_burden_2022}, that could shed light on the usage of more words across the dimensions of in-group polysemy and between-groups divergence. We also experimented with using one of the newest generative LLMs as a data annotator, with promising results of agreement with human annotators that does not fall far behind their inter-rater (dis)agreement. While the results of machine learning models (including LLMs) should be always be critically evaluated, we are reaching a point where they could be used in lieu of human annotators on larger, more tedious or costly tasks, where if the task is simple enough (which can, again, be evaluated using smaller human test sets).

The dataset, consisting of written American English as used on a micro-blogging platform, of course has its limitations, including questions like how generalizable and representative of the given society and language it may be --- in this case the United States, and US American English.
Naturally, the demographics of users of an online platform like Twitter may not be representative of the society as a whole.
A number of previous studies on political differences and polarization cited above have focused on high-profile personas such as politicians or influential opinion leaders, using their writings, speeches or social media content as data. 
We were interested in unedited natural language as used by regular people in everyday situations. Such naturalistic data is, however, hard to acquire in large volumes from offline usage --- but relatively easy to mine from social media. We accept that language usage on Twitter may only represent a part of the linguistic repertoire and competence of speakers of a given language, and online language use as such may be situational and differ from person-to-person communication \parencite[cf.][]{mcculloch_because_2019,joseph_misalignment_2021}. 
However, a case could be made that the only way to observe natural language data is inevitably to observe it in some variety or other; in our case it just happens to be the online one.

\subsection{Future research}

Data mining Twitter/X, while popular until recently in fields like computational social science, has become difficult given shifts in the platform's policies \parencite{ledford_researchers_2023}. 
As outlined in the Methods however, the proposed framework is in principle applicable to any platform or network where users and links between users can be identified and the data collected. 
Social media examples may include Reddit or other forum-type platforms (user groups could be grouped by subreddits or subforums they do and do not subscribe to), Wikipedia (e.g. editors grouped by domains where they do and do not edit), or any of the Twitter-like platforms that have emerged following the rebranding and other changes in Twitter, if their policies and infrastructure enables academic research.

Follow-up work could look into aspects of potential differences across political divides other than just lexico-statistics and lexical semantics. While easily inferred from textual social media data, these are by no means the only avenues of variation in language. 
We focused on text, but it may be interesting to compare visual media like profile pictures \parencite[as a form of self-representation;][]{kapidzic_race_2015,robertson_emoji_2020}, and posted images, memes or videos \parencite[scalable using machine learning just as textual data; cf.][]{verma_sentiment_2020,beskow_evolution_2020}.

More broadly, the same operational logic could be used to study other cultural and social domains where more or less complete user or participant data is available. 
For example, \textcite{zemaityte_quantifying_2023} investigate a large dataset from a globally-used platform of film professionals and film festivals; the same approach could be used to delineate potentially diverging groups such as filmmakers (by which festivals they frequent and which they do not).
Similarly, television production crews and groupings of individuals and the content they produce could be studied where complete production or historical databases are available
\parencite[cf.][]{ibrus_quantifying_2022-1,oiva_framework_2023}.

Finally, in the linguistic domain, it would be of particular interest to disentangle the relationships between the variation observed along political affiliations and the different sources of underlying natural variation \parencite[e.g. regional, as explored in the US context by][]{louf_american_2023}, eventually both in varieties of English and other languages. If this would be possible, then it could be determined if some of the variation or divergence we observe here could be purely politically driven --- as in, not an effect of regional or social differences, but use of in-group markers to express political leanings \parencite[cf.][]{albertson_dog-whistle_2015}.

\section{Conclusions}

We proposed an approach to delineate groups of users on social media according to their interaction statistics on a given platform, mined a large corpus of US American English language tweets from Twitter, and used a combination of machine learning, lexico-statistical, and human data annotation methods to estimate and illustrate the extent of lexical and semantic differences in the language use of the left-leaning and right-leaning polarities in the US.
While we focused here on a single language and social media platform, we hope the general framework to be a useful contribution for data-driven computational research into language variation and change more generally.

\section*{Acknowledgments and author contributions}

A.K. designed the study, collected the data, conducted data analysis and data annotation, created the figures, and wrote the manuscript. C.C. also designed the study, conducted data annotation, and wrote the manuscript.
A.K. is supported by the CUDAN ERA Chair project for Cultural Data Analytics, funded through the European Union Horizon 2020 research and innovation program (Project No. 810961).
C.C. was supported during a part of this research by the ESRC Research Grant "Constraints on the adaptiveness of information content in language (CAIL): Improving communication and detecting failures in audience design" (ES/T005955/1).

\section*{Data and code availability}

The code used to run the analyses is available at \\ 
\url{https://github.com/andreskarjus/evolving_divergence} \\
Unfortunately, and exceptionally, at this time we cannot make neither the collected data nor the tweet or user IDs publicly available, in order to avoid potential conflicts with the current Terms of Service of the Twitter/X platform regarding potentially political and sensitive contexts. The data may be shared directly upon reasonable request.

\FloatBarrier
\begingroup
\setlength{\emergencystretch}{8em}
\printbibliography
\endgroup
\FloatBarrier

\newpage
\section*{Appendix}
\setcounter{table}{0}
\renewcommand{\table}{A\arabic{table}}

This is a list of the news accounts we collected from AllSides and mined for user listings on Twitter. The data was current as of the start of the data collection period in mid-2021.

\centering
\begin{longtable}{lllr}
  \hline
Account name & Account user name & AllSides designation & Followers counted \\ 
  \hline
MSNBC Daily & msnbcdaily & left & 178983 \\ 
  MSNBC & msnbc & left & 4323778 \\ 
  New York Times Opinion & nytopinion & left & 794242 \\ 
  Democracy Now! & democracynow & left & 798840 \\ 
  AlterNet & alternet & left & 136033 \\ 
  HuffPost & huffpost & left & 11355704 \\ 
  HuffPost Politics & huffpostpol & left & 1441594 \\ 
  Jacobin & jacobin & left & 339512 \\ 
  Mother Jones & motherjones & left & 860441 \\ 
  BuzzFeed & buzzfeed & left & 6340783 \\ 
  BuzzFeed News & buzzfeednews & left & 1385389 \\ 
  The Daily Beast & thedailybeast & left & 1333875 \\ 
  The Intercept & theintercept & left & 865950 \\ 
  Slate & slate & left & 1778853 \\ 
  Vox & voxdotcom & left & 1027060 \\ 
  The Nation & thenation & left & 1273594 \\ 
  The New Yorker & newyorker & left & 8967134 \\ 
  CNN & cnn & left & 54573337 \\ 
  CNN Breaking News & cnnbrk & left & 61332975 \\ 
  CNN Newsroom & cnnnewsroom & left & 459580 \\ 
  CNN Politics & cnnpolitics & left & 4179383 \\ 
  ABC News & abc & leanleft & 16807969 \\ 
  ABC News Politics & abcpolitics & leanleft & 999654 \\ 
  AP Politics & ap\_politics & leanleft & 561001 \\ 
  The Atlantic & theatlantic & leanleft & 2112540 \\ 
  The New York Times & nytimes & leanleft & 50435924 \\ 
  Bloomberg Politics & bpolitics & leanleft & 342832 \\ 
  Bloomberg & business & leanleft & 7496915 \\ 
  CBS & cbs & leanleft & 1176160 \\ 
  The Economist & theeconomist & leanleft & 25796060 \\ 
  Guardian US & guardianus & leanleft & 207944 \\ 
  NBS Television & nbstv & leanleft & 916940 \\ 
  NPR & npr & leanleft & 8752334 \\ 
  NPR Politics & nprpolitics & leanleft & 3042719 \\ 
  POLITICO & politico & leanleft & 4515468 \\ 
  ProPublica & propublica & leanleft & 927537 \\ 
  TIME & time & leanleft & 18255872 \\ 
  The Washington Post & washingtonpost & leanleft & 18225814 \\ 
  Yahoo News & yahoonews & leanleft & 1092791 \\ 
  The Associated Press & ap & center & 15295288 \\ 
  Axios & axios & center & 596867 \\ 
  The Christian Science Monitor & csmonitor & center & 80401 \\ 
  IJR & theijr & center & 104467 \\ 
  MarketWatch & marketwatch & center & 4072708 \\ 
  Newsweek & newsweek & center & 3446005 \\ 
  Reuters & reuters & center & 23748322 \\ 
  RealClearPolitics & realclearnews & center & 212363 \\ 
  The Hill & thehill & center & 4243021 \\ 
  USA TODAY & usatoday & center & 4426356 \\ 
  The Wall Street Journal & wsj & center & 19001174 \\ 
  WSJ Editorial Page & wsjopinion & leanright & 112390 \\ 
  The American Conservative & amconmag & leanright & 55513 \\ 
  Fox News & foxnews & leanright & 20192707 \\ 
  Fox News Alert & foxnewsalert & leanright & 497512 \\ 
  The Dispatch & thedispatch & leanright & 38606 \\ 
  The Epoch Times & epochtimes & leanright & 424314 \\ 
  New York Post & nypost & leanright & 2270594 \\ 
  Washington Examiner & dcexaminer & leanright & 287104 \\ 
  The Washington Times & washtimes & leanright & 414124 \\ 
  Newsmax & newsmax & right & 962844 \\ 
  FOX \& friends & foxandfriends & right & 1207044 \\ 
  Fox News Opinion & foxnewsopinion & right & 137908 \\ 
  Breitbart News & breitbartnews & right & 1464601 \\ 
  The Federalist & fdrlst & right & 279956 \\ 
  One America News & oann & right & 1489997 \\ 
  Daily Caller & dailycaller & right & 805087 \\ 
  Daily Mail US & dailymail & right & 356926 \\ 
  CBN News & cbnnews & right & 150614 \\ 
  TheBlaze & theblaze & right & 680715 \\ 
  The Daily Wire & realdailywire & right & 652916 \\ 
  National Review & nro & right & 329127 \\ 
  The American Spectator & amspectator & right & 40051 \\ 
   \hline
\end{longtable}

\end{document}